\documentstyle[sprocl]{article}

\input{psfig}

\def\be{\begin{equation}}
\def\ee{\end{equation}}
\def\bea{\begin{eqnarray}}
\def\eea{\end{eqnarray}}
\begin{document}

\title{AN ALGEBRAIC MODEL OF BARYON SPECTROSCOPY}

\author{R. BIJKER}

\address{ICN-UNAM, A.P. 70-543, 04510 M\'exico, D.F., M\'exico} 

\author{A. LEVIATAN}

\address{Racah Institute of Physics, The Hebrew University,
Jerusalem 91904, Israel}

\maketitle

\abstracts{We discuss recent calculations of the mass spectrum,  
electromagnetic and strong couplings of baryon resonances. 
The calculations are done in a collective constituent model 
for the nucleon, in which the resonances are interpreted as 
rotations and vibrations of a symmetric top with a prescribed 
distribution of the charge and magnetization. 
We analyze recent data on eta-photo- and eta-electroproduction, 
and the tensor analyzing power in deuteron scattering.} 

\section{Introduction}

Effective models of baryons based on three constituents
share a common spin-flavor-color structure but differ in their 
treatment of the spatial dynamics. 
Quark potential models in nonrelativistic \cite{nrqm} or relativized 
\cite{rqm} forms emphasize the single-particle aspects of quark dynamics 
for which only a few low-lying configurations in the confining potential 
contribute significantly to the eigenstates of the Hamiltonian. 
On the other hand, some regularities
in the observed spectra ({\it e.g.} linear Regge trajectories,
parity doubling) hint that an alternative, collective
type of dynamics may play a role in the structure of baryons. 

In this contribution we present a collective model of 
baryons within the context of an algebraic approach \cite{BIL} 
in which the baryon resonances are interpreted as rotations and 
vibrations of a symmetric top. We analyze the mass spectrum,  
the electromagnetic and strong couplings of nonstrange baryons. 
Particular attention is paid to the new values of the helicity 
amplitudes of the N(1520)$D_{13}$ and N(1535)$S_{11}$ resonances  
and their model independent ratios 
\cite{Krusche,Nimai1,Nimai2,Tiator,Armstrong}, the tensor 
analyzing power in deuteron scattering \cite{Tomasi}, and the 
strong decays of the N(1535) resonance.   

\section{Algebraic model}

We consider baryons to be built of three constituent parts 
which are characterized by both internal and radial (or spatial) 
degrees of freedom. 
The internal degrees of freedom of these three parts are taken to be:
flavor-triplet $u,d,s$ (for the light quark flavors), 
spin-doublet $S=1/2$, and color-triplet. The internal algebraic 
structure of the constituent parts consists of the usual spin-flavor 
(${\rm sf}$) and color (${\rm c}$) algebras 
$SU_{\rm sf}(6) \otimes SU_{\rm c}(3)$. 
The difference between models of baryon spectroscopy lies in the 
treatment of the spatial dynamics. The relevant radial degrees of freedom 
for the relative motion of the three constituent parts of this  
configuration can be taken as the Jacobi coordinates. These can be 
treated algebraically in terms of the spectrum generating algebra of $U(7)$ 
for the radial (or orbital) excitations \cite{BIL}. The model space 
is spanned by the symmetric irreducible representation $[N]$ of $U(7)$, 
and contains the harmonic oscillator shells with
$n=n_{\rho}+n_{\lambda}=0,1,2,\ldots, N$. The value of $N$ determines
the size of the model space and, in view of confinement, is expected 
to be large. 
The full algebraic structure is obtained by combining the radial 
part with the internal spin-flavor-color part 
\bea
{\cal G} \;=\; U(7) \otimes SU_{\rm sf}(6) \otimes SU_{\rm c}(3) ~.
\eea
The radial part of the baryon wave function has to be combined 
with the spin-flavor and color parts, in such a way that the total 
wave function is antisymmetric. 

A convenient set of basis states for the baryon wave functions 
is provided by the case of three 
identical constituents, for which the radial and spin-flavor parts 
of the baryon wave function are in addition labeled by their 
transformation properties under the permutation group $S_3$: 
$t=S$ for the symmetric, $t=A$ for the antisymmetric and $t=M$ 
mixed symmetry representation. The wave functions can now be labeled as 
\bea
\left| \, ^{2S+1}\mbox{dim}\{SU_{\rm f}(3)\}_J \,
[\mbox{dim}\{SU_{\rm sf}(6)\},L^P],(v_1,v_2);K \, \right> ~. \label{wf}
\eea
The spin-flavor part is characterized by the dimensions of the 
irreducible representations of  
$SU_{\rm sf}(6) \supset SU_{\rm f}(3) \otimes SU_{\rm s}(2)$. 
For the radial part we consider a collective (string-like) model 
in which the baryons are interpreted as exitations of an oblate 
top \cite{BIL}. In this case the spatial part is characterized by 
the labels: $(v_1,v_2);K,L^P_t$, where $(v_1,v_2)$ 
denotes the vibrations (stretching and bending) of the oblate top; 
$K$ denotes the projection of the 
rotational angular momentum $L$ on the body-fixed symmetry-axis; 
$P$ the parity and $t$ the $S_3$ symmetry type of the state under 
permutations. The permutation symmetry of the spatial part must be the
same as that of the spin-flavor part to ensure the antisymmetry 
of the total baryon wave-function: $t=S \leftrightarrow [56]$, 
$t=A \leftrightarrow [20]$ and $t=M \leftrightarrow [70]$.  
The spin $S$ and the orbital angular momentum $L$ are coupled to total 
angular momentum $J$.

\section{Mass spectrum} 

The mass operator depends both on the radial and the internal 
degrees of freedom. The masses of nonstrange baryons have been 
analyzed with the mass formula \cite{BIL} 
\bea
M^2 &=& M^2_0 + \kappa_1 \, v_1 + \kappa_2 \, v_2 + \alpha \, L 
+ a \, \Bigl[ \langle \hat C_2(SU_{ \rm sf}(6)) \rangle - 45 \Bigr]
\nonumber\\
&& + b \, \Bigl[ \langle \hat C_2(SU_{\rm f}(3)) \rangle -  9 \Bigr]
   + c \, \Bigl[ S(S+1) - \frac{3}{4} \Bigr] ~.
\label{massformula}
\eea
For the radial part we have adopted a collective model of the 
nucleon in which the baryons are interpreted as rotational 
and vibrational excitations of an oblate symmetric top. 
The spectrum consists of a series of vibrational excitations 
labeled by $(v_1,v_2)$, and a tower of rotational excitations built 
on top of each vibration. The occurrence of linear Regge trajectories 
suggests to add, in addition to the vibrational frequencies 
$\kappa_1$ and $\kappa_2$, a term linear in $L$. The slope of these 
trajectories is given by $\alpha$. 
For the spin-flavor part of the mass operator we use a 
G\"ursey-Radicati \cite{GR} form. 
Interaction terms that mix the space and internal degrees 
of freedom such as for example spin-orbit and tensor couplings 
have not been included. 

The coefficients are determined in a simultaneous fit 
to all well-established (three and four star \cite{PDG})  
nucleon and $\Delta$ resonances. 
The N$(1440)P_{11}$ and $\Delta(1600)P_{33}$ resonances are 
assigned to the $(v_1,v_2)=(1,0)$ vibration and the 
N$(1710)P_{11}$ resonance to the $(v_1,v_2)=(0,1)$ vibration, 
whereas the negative parity resonances N$(1520)D_{13}$, 
N$(1535)S_{11}$ and N$(1650)S_{11}$ resonances are interpreted 
as rotational excitations. 
We find a good overall fit for 24 resonances with a r.m.s. 
deviation of $\delta_{\mbox{rms}}=39$ MeV \cite{BIL}. 

A common feature to all $q^3$ quark models is the occurrence of 
missing resonances. 
In a recent three-channel analysis evidence was found for the 
existence of a $P_{11}$ nucleon resonance at $1740 \pm 11$ MeV 
\cite{Zagreb}. It is tempting to assign this resonance as one 
of the missing resonances \cite{CLRS}. In the present calculation 
it is associated with the $^{2}8[20,1^+]$ configuration and 
appears at 1720 MeV, compared to 1880 MeV in the relativized 
quark model \cite{rqm}.   

\section{Collective form factors}

In addition to the masses, it is important to study 
decay processes which are far more sentistive to details 
in the baryon wave functions. Under the assumption that 
the electromagnetic (strong) decays involve the coupling 
of a photon (elementary meson) to a single constituent, 
the transition operators that induce these decay processes 
can be expressed in terms of the algebraic operators \cite{BIL}
\bea
\hat U &=& \mbox{e}^{i k \beta \hat D_{\lambda,z}/X_D} ~,
\nonumber\\
\hat T_m &=& - \frac{i m_3 k_0 \beta}{2X_D} \left(
\hat D_{\lambda,m} \, \mbox{e}^{i k \beta \hat D_{\lambda,z}/X_D} +
\mbox{e}^{i k \beta \hat D_{\lambda,z}/X_D} \,
\hat D_{\lambda,m} \right) ~. 
\label{ut2}
\eea
Here $(k_0,\vec{k})$ is the four-momentum of the emitted quantum.
The dipole operator $\hat D_{\lambda,m}$ is a generator of $U(7)$ and
$X_D$ its normalization; $\beta$ is a radial coordinate 
\cite{BIL,emff}. Different types of collective models are specified by
a distribution of the charge and magnetization along the string 
$g(\beta)$. The collective form factors are then obtained
by folding the matrix elements of $\hat U$ and 
$\hat T_{m}$ with this probability distribution 
\bea
{\cal F}(k)   &=& \int \mbox{d} \beta \, g(\beta) \,
\langle \psi_f | \hat U   | \psi_i \rangle ~,
\nonumber\\
{\cal G}_m(k) &=& \int \mbox{d} \beta \, g(\beta) \,
\langle \psi_f | \hat T_{m} | \psi_i \rangle  ~.
\label{radint}
\eea
Here $\psi$ denotes the spatial part of the baryon wave function.
We use the ansatz 
\bea
g(\beta) &=& \beta^2 \, \mbox{e}^{-\beta/a}/2a^3 ~, \label{gbeta}
\eea
to obtain the dipole form for the elastic form factor.
The coefficient $a$ is a scale parameter.
With the same distribution we can now derive closed expressions 
for the inelastic or transition form factors \cite{BIL,emff}. 
These collective form factors drop as powers of $k$. 
This property is well-known experimentally and is 
in contrast with harmonic oscillator quark models in which all 
form factors which fall off exponentially. 

All helicity amplitudes, form factors and decay widths for 
electromagnetic (transverse, longitudinal and scalar) and strong 
couplings can be expressed in terms of the radial matrix 
elements ${\cal F}$ and ${\cal G}_m$, 
a spin-flavor matrix element and a phase space factor. 

\section{Electromagnetic couplings}

In constituent models, electromagnetic couplings arise from the 
coupling of the constituent parts to the electromagnetic 
field. We discuss here the case of the emission of a lefthanded photon 
from a single constituent 
\bea
B \rightarrow B' + \gamma ~, 
\eea
for which the nonrelativistic part of the electromagnetic coupling 
is given by 
\bea
{\cal H}_{em} &=& 6 \sqrt{\frac{\pi}{k_0}} \mu_3 e_3 \,  
\left[ k s_{3,-} \hat U - \frac{1}{g_3} \hat T_{-} \right] ~, 
\label{hem} 
\eea
where $e_3$, $\mu_3$, $g_3$ and $s_3$ are the charge, scale magnetic 
moment, $g$ factor and spin of the third consituent. 
In order to avoid ambiguities 
arising from the treatment of recoil effects, calculations are 
carried out in the equal momentum or Breit frame. 
The scale parameter $a$ of Eq.~(\ref{gbeta}) was determined from 
a simultaneous fit to the proton and neutron charge radii, and the 
electric and magnetic form factors of the proton and neutron 
to be $a=0.232$ fm \cite{emff}. 

\subsection{Ratios of photocouplings}

\begin{table}
\centering
\caption[]{Ratios of helicity amplitudes}
\label{ratios} 
\vspace{15pt} 
\small
\begin{tabular}{|lcc|}
\hline
& & \\
& N(1535)$S_{11}$ & N(1520)$D_{13}$ \\
& $A^n_{1/2}/A^p_{1/2}$ & $A^p_{3/2}/A^p_{1/2}$ \\
& & \\
\hline
& & \\
Feynman et al.   \cite{FKR}      & $-0.69$ & $-3.21$ \\
Koniuk and Isgur \cite{KI}       & $-0.81$ & $-5.57$ \\
Warns et al.     \cite{Warns2}   & $-1.06$ & $-9.00$ \\
Close and Li     \cite{CL}       & $-0.74$ & $-6.55$ \\
                                 & $-0.65$ & $-4.87$ \\
Li and Close     \cite{LC}       & $-0.54$ & $-2.50$ \\
                                 & $-0.56$ & $-2.61$ \\
Capstick         \cite{Capstick} & $-0.83$ & $-8.93$ \\
Bijker et al.    \cite{BIL}      & $-0.69$ & $-2.53$ \\
                                 & $-0.81^a$ & \\
Santopinto et al. \cite{SIG}     & $-0.68$ & $-1.55$ \\
                                 & $-0.67$ & $-1.80$ \\
& & \\
Mukhopadhyay et al. \cite{Nimai1,Nimai2}  
& $-0.84 \pm 0.15$ & $-2.5 \pm 0.2 \pm 0.4$ \\
Tiator et al. \cite{Tiator} & & $-2.1 \pm 0.2$ \\
PDG \cite{PDG}      & $-0.51 \pm 0.34$ & $-6.9 \pm 2.6$ \\
& & \\
\multicolumn{3}{|l|}{$^a$ Calculated with a mixing angle 
of $\theta=-38^{\circ}$} \\
& & \\
\hline
\end{tabular}
\normalsize
\end{table}

Recent experiments on eta-photoproduction have  
yielded valuable new information on the helicity amplitudes of the 
N(1520)$D_{13}$ and N(1535)$S_{11}$ resonances. 
In an Effective Lagrangian Approach model-independent ratios 
of photocouplings were extracted from the new data 
\cite{Nimai1,Nimai2} 
\bea
\mbox{N(1535)}S_{11} &: \hspace{1cm}& 
A^n_{1/2}/A^p_{1/2} \;=\; -0.84 \pm 0.15 
\nonumber\\
\mbox{N(1520)}D_{13} &: \hspace{1cm}& 
A^p_{3/2}/A^p_{1/2} \;=\; -2.5 \pm 0.2 \pm 0.4 
\eea
The latter value was confirmed in a recent analysis by Tiator et al. 
\cite{Tiator} who found $-2.1 \pm 0.2$~. 
These values are in good agreement with those of the collective 
algebraic model, $-0.69$ and $-2.53$, respectively \cite{BIL}. 

In the analysis of the photocouplings of the N$(1535)S_{11}$ and 
N$(1650)S_{11}$ resonances it was found that these resonances 
are mixtures of the $|^{2}8_{1/2}[70,1^-] \rangle$ and 
$|^{4}8_{1/2}[70,1^-] \rangle$ configurations \cite{FH}. 
The transition to the second configuration 
is forbidden by the Moorhouse selection rule \cite{Moorhouse}. 
This mixing breaks $SU_{\rm sf}(6)$ and can only be introduced 
by a tensor-like coupling. It appears to be the only place in 
the spectrum and transitions where there is a clear evidence
of such a breaking. With a mixing angle of $\theta=-38^{\circ}$ 
\cite{BIL} the ratio of photocouplings of the 
N(1535)$S_{11}$ resonance increases from $-0.69$ to $-0.81$ 
\cite{BIL}. 

In Table~\ref{ratios} we compare these values with some model 
calculations. Whereas for the ratio of the photocouplings 
of the N(1535)$S_{11}$ resonance there is relatively 
little variation between the various theoretical results, 
for the ratio of the proton photocouplings of 
the N(1520)$D_{13}$ resonance there is a large spread in values. 
We also note the large discrepancy between the photocouplings obtained 
from pion-photoproduction \cite{PDG} and the new values 
determined from eta-photoproduction \cite{Nimai2,Tiator}.  

\subsection{Transition form factors}

\begin{figure}
\centerline{\psfig{file=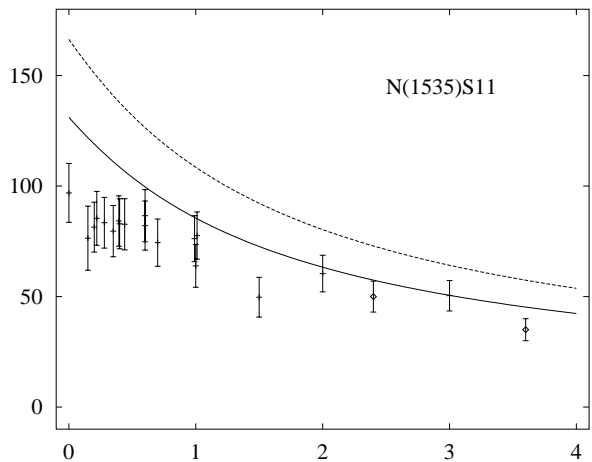,width=0.75\linewidth}}
\vspace{15pt}
\caption[]{N(1535)$S_{11}$ proton helicity amplitude in 
$10^{-3}$ GeV$^{-1/2}$ as a function of $Q^2$ in (GeV/c)$^2$. 
A factor of $+i$ is suppressed. The data are taken from 
a compilation in \protect\cite{Armstrong}. The solid and 
dashed curves correspond to mixing angles $\theta=-38^{\circ}$ 
and $0^{\circ}$, respectively.}
\label{n1535}
\end{figure}

There is currently much interest in the $Q^2$ dependence of the 
transition form factors. 
In Fig.~\ref{n1535} we show the N(1535)$S_{11}$ proton helicity 
amplitude $A^p_{1/2}$ for which there exist 
interesting new data \cite{Armstrong} (diamonds). The other points 
are obtained from a reanalysis of old(er) data, but now using the same 
values of the resonance parameters for all cases. 
The solid and dashed 
curves represent the results of the collective model which 
were obtained by introducing a mixing angle of 
$\theta = -38^{\circ}$ and $0^{\circ}$, respectively. 
Just as for the photocouplings, the introduction of the mixing 
angle improves the agreement with the data. 

\subsection{Tensor analyzing power}

\begin{figure}
\centerline{\psfig{file=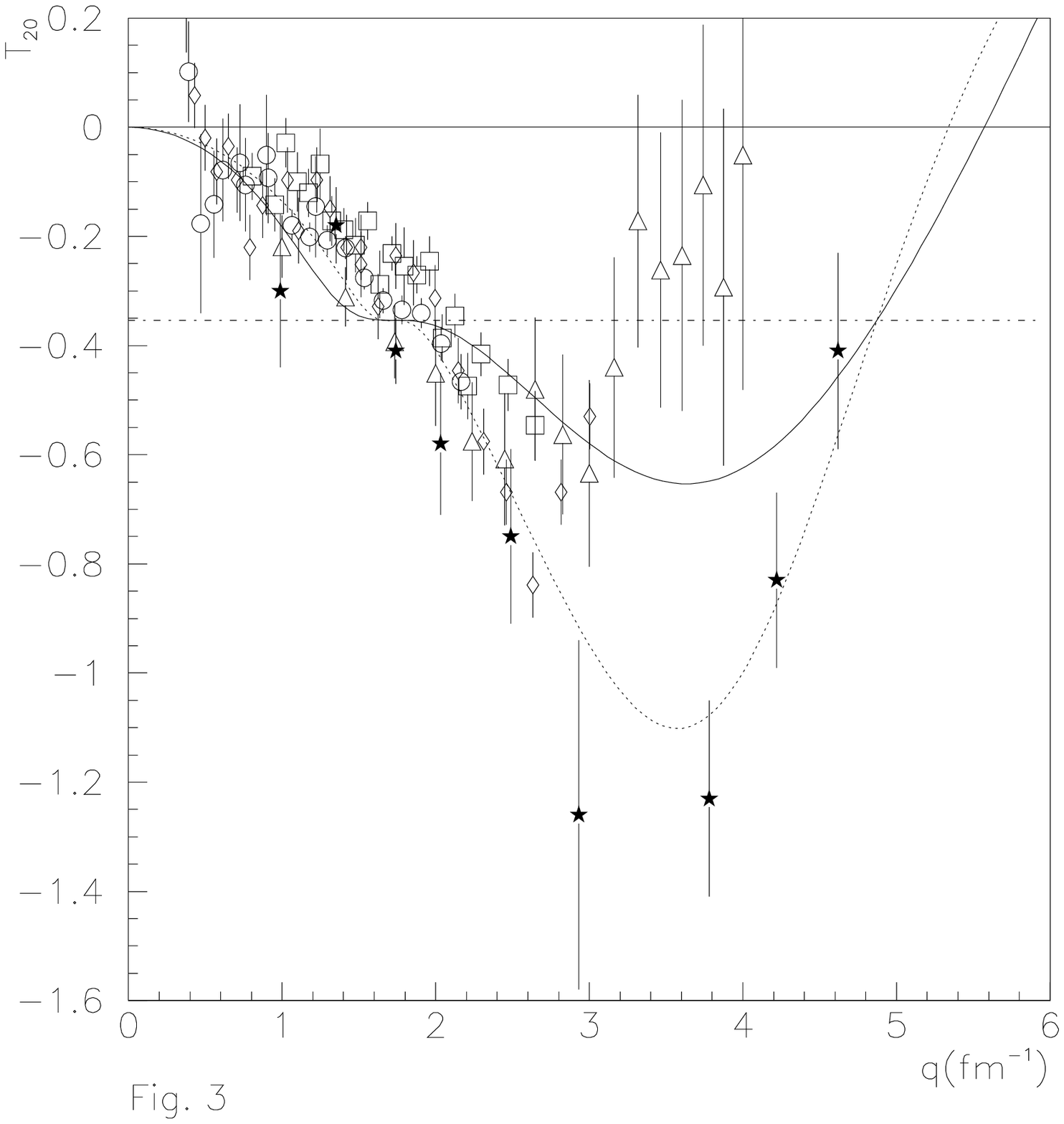,width=0.5\linewidth}}
\vspace{-1cm}
\caption[]{Tensor analyzing power $T_{20}$ for
$e^-+d \rightarrow e^-+d$ elastic scattering (filled stars) and 
$d+p \rightarrow d+X$ at incident momenta of 3.75 GeV/c
(open diamonds), 5.5 GeV/c (open circles),
4.5 GeV/c (open squares), 9 GeV/c (open triangles). 
The data are taken from a compilation in 
\protect\cite{Tomasi}. 
The curves represent the results in an $\omega-$exchange 
model with $r=0$ (dashed-dotted line), and $r$ calculated 
using collective form factors for the Roper resonance only 
(dotted line) and for the Roper, N(1520),  
N(1535) and N(1650) resonances (solid line).}
\label{t20}
\end{figure}

Recently, we studied the tensor analyzing power $T_{02}$ of the 
process $d+p \rightarrow d+X$ for forward deuteron scattering in 
the framework of $\omega$ exchange using the collective model 
for the nucleon resonances \cite{Tomasi}. 
Polarization variables, such as the tensor analyzing power, 
are sensitive to the ratio $r = \sigma_L/\sigma_T$ of the 
cross sections for the absorption of virtual isoscalar photons 
with longitudinal and transversal polarizations by nucleons 
\cite{Rekalo}. Since the lowlying negative parity resonances 
N(1520)$D_{13}$, N(1535)$S_{11}$ and N(1650)$S_{11}$ have only 
isovector longitudinal form factors, the tensor analyzing power 
is especially sensitive to the nonvanishing isoscalar longitudinal 
form factor of the N(1440)$P_{11}$ Roper resonance. 
Without the excitation of the Roper resonance $r=0$, and
the value of $T_{20}$ becomes a constant: $T_{20}=-1/\sqrt{8}$, 
(dashed-dottel line in Fig.~\ref{t20}) in disagreement with 
the existing data (open symbols in Fig~\ref{t20}). 
The dotted line includes only the Roper resonance, 
whereas the solid line includes the negative parity resonances 
N(1520)$D_{13}$, N(1535)$S_{11}$ and N(1650)$S_{11}$ as well. 
Fig.~\ref{t20} shows that a good description of the data is 
obtained when $r$ is calculated using collective form factors. 

\section{Strong couplings}

\begin{table}
\centering 
\caption[]{$N^{\ast} \rightarrow N + \pi$ and 
$N^{\ast} \rightarrow N + \eta$ 
decay widths of 3* and 4* resonances in MeV. 
n.o. stands for not observed. 
The experimental values are taken from \protect\cite{PDG}.}
\label{nstar}
\vspace{15pt}
\small
\begin{tabular}{|lcccc|}
\hline
& & & & \\
Baryon & \multicolumn{2}{c}{$\Gamma(N^{\ast} \rightarrow N \pi)$} 
& \multicolumn{2}{c|}{$\Gamma(N^{\ast} \rightarrow  N \eta)$} \\
& th & exp & th & exp \\
& & & & \\
\hline
& & & & \\
N(1520)$D_{13}$   & $115$ & $ 67 \pm  9$ & $ 1$ & n.o. \\
N(1535)$S_{11}$   & $ 85$ & $ 79 \pm 38$ & $ 0$ & $74 \pm 39$ \\
                  & $ 20^a$ &            & $ 0^a$ & \\
N(1650)$S_{11}$   & $ 35$ & $130 \pm 27$ & $ 8$ & $11 \pm  6$ \\
                  & $142^a$ &            & $ 0^a$ & \\
N(1675)$D_{15}$   & $ 31$ & $ 72 \pm 12$ & $17$ & n.o. \\
N(1680)$F_{15}$   & $ 41$ & $ 84 \pm  9$ & $ 1$ & n.o. \\
N(1700)$D_{13}$   & $  5$ & $ 10 \pm  7$ & $ 4$ & n.o. \\
N(1720)$P_{13}$   & $ 31$ & $ 22 \pm 11$ & $ 0$ & n.o. \\
N(2190)$G_{17}$   & $ 34$ & $ 67 \pm 27$ & $11$ & n.o. \\
N(2220)$H_{19}$   & $ 15$ & $ 65 \pm 28$ & $ 1$ & n.o. \\
N(2250)$G_{19}$   & $  7$ & $ 38 \pm 21$ & $ 9$ & n.o. \\
N(2600)$I_{1,11}$ & $  9$ & $ 49 \pm 20$ & $ 3$ & n.o. \\
& & & & \\
\multicolumn{5}{|l|}{$^a$ Calculated with a mixing angle 
of $\theta=-38^{\circ}$} \\
& & & & \\
\hline
\end{tabular}
\normalsize
\end{table}

Next we consider strong decays of baryons by the emission of
a pseudoscalar meson
\bea
B \rightarrow B^{\prime} + M ~.
\eea
Here we use an elementary emission model in which the meson 
is emitted from a single constituent \cite{KI,strong} 
\bea
{\cal H}_s &=& \frac{1}{(2\pi)^{3/2} (2k_0)^{1/2}} \, 6 X^M_{3} 
\nonumber\\
&& \hspace{1cm} 
\left[ (gk-\frac{1}{6}hk) s_{3,z} \hat U - h s_{3,z} \hat T_z 
- \frac{1}{2} h (s_{3,+} \hat T_- + s_{3,-} \hat T_+) \right] ~. 
\label{hstrong}
\eea
The flavor operator $X^M_3$ corresponds to the emission of an
elementary meson by the third constituent. 
For the pseudoscalar $\eta$ mesons we introduce a mixing angle 
$\theta_P=-23^{\circ}$ between the octet and singlet mesons 
\cite{PDG,GIK}. The strong decay widths are calculated in the 
rest frame of the decaying resonance \cite{LeYaouanc}. 

We consider strong decays of nonstrange baryons into 
the $\pi$ and $\eta$ channels \cite{strong}. 
The decay widths depend on the parameters $g$ and $h$ 
in Eq.~(\ref{hstrong}) and on the scale parameter $a$ of 
Eq.~(\ref{gbeta}). These parameters were determined 
from a least square fit to the $N \pi$ partial widths 
(which are relatively well known) with the exclusion
of the N(1535)$S_{11}$ and N(1650)$S_{11}$ resonances. 
For the latter, the situation is not clear due to possible 
mixing of different $S_{11}$ states or even the 
existence of a third $S_{11}$ resonance \cite{LW}. As a
result we find $g=1.164$ GeV$^{-1}$ and $h=-0.094$ GeV$^{-1}$. 
The relative sign is consistent with a previous analysis of the strong
decay of mesons \cite{GIK} and with a derivation from the axial-vector
coupling (see \cite{LeYaouanc}). The value of the scale parameter 
$a=0.232$ fm is found to be equal to the value extracted in the 
calculation of the electromagnetic couplings \cite{emff}.
The coefficients $g$, $h$ and $a$ are kept equal for {\em all} 
resonances and {\em all} decay channels. 

In Table~\ref{nstar} we show the decay widths of the nucleon 
resonances into the $N \pi$ and $N \eta$ channels. 
The $N \pi$ decay widths are found to be in fair agreement 
with experiment. For the $N \eta$ channel our calculation gives  
systematically small values, whereas the PDG compilation \cite{PDG} 
lists a large $\eta$ width for the N$(1535)S_{11}$ resonance. 
We emphasize that the coefficients $g$ and $h$ in the transition 
operator were determined from the $N + \pi$ decays, so that 
the $\eta$ decays are
calculated without introducing any further parameters.
The $\eta$ decays are suppressed relative to the $\pi$ 
decays because of phase space factors. 

The results of our analysis suggest that the large $\eta$ width 
that is usually attributed to the N(1535)$S_{11}$ resonance is 
not due to a conventional $q^3$ state. One possible explanation 
is the presence of another state in the same mass region,
{\it e.g.} a quasi-bound meson-baryon $S$ wave resonance just below
or above threshold, for example $N\eta$, $K\Sigma$ or $K\Lambda$
\cite{Kaiser}. Another possibility is an exotic configuration of 
four quarks and one antiquark ($q^{4}\bar{q}$). 

\section{Summary and conclusions}

In this contribution we have analyzed the masses, electromagnetic 
and strong couplings of baryon resonances in a collective model 
of the nucleon, in which the baryons are interpreted as 
rotations and vibrations of a symmetric top with a prescribed 
distribution of the charge and magnetization. 

A study of the mass spectrum showed that the third of four $P_{11}$ 
resonances in the recent Zagreb analysis \cite{Zagreb} can be assigned 
as the lowest missing resonance that arises in the collective model. 
Far more sensitive tests of baryon models are provided by the 
electromagnetic and strong couplings. As an example we studied 
the recently determined ratios of photocouplings of the N(1535) and 
N(1520) resonances. Especially for the N(1520) resonance there is a large 
variation in the model predictions. Both ratios are found to be in 
good agreement with the predicted values of the collective model. 
The same conclusion holds for the transition form factor of the 
N(1535) resonance. Another example is the tensor analyzing power in 
deuteron scattering which is very 
sensitive to the isoscalar longitudinal form factor of the 
Roper resonance. It was shown that the use of collective form factors 
gives the correct dependence of $T_{20}$ on the momentum transfer.

Whereas the helicity amplitudes for photo- and 
electroproduction of the N(1535) resonance are 
described well, there is a large discrepancy for the strong couplings. 
Our analysis of the strong decay widths shows that while the 
$\pi$ decays follow the expected pattern, the decays into 
$\eta$ exhibit some unusual features. Our calculations do not 
show any indication for a large $\eta$ width, as is observed 
for the N(1535) resonance. This seems to indicate the 
presence of configurations other than $q^3$ in the same mass 
region. 

The nature of the N(1535) resonance has been addressed recently by 
many authors (see {\em e.g.} \cite{Tiator,Kaiser,Hohler,Feuster}). 
For example, in the chiral meson-baryon Lagrangian framework of 
\cite{Kaiser} it was found that the N(1535) `could well be a strong 
background instead of a clean resonance'; in \cite{Hohler} it was shown 
that speed plots show no structure in the $S_{11}$ partial wave 
at 1535 MeV, but only the strong $N\eta$ cusp and a resonance 
at 1650 MeV. 

In conclusion, we have shown that the collective model of 
baryons provides a good overall description of the available 
data. The nature of the N(1535) resonance remains an open 
and intriguing question. 

\section*{Acknowledgements}

This work is supported in part by DGAPA-UNAM under
project IN101997.

\end{document}